Arif Raza, Zaka-ul-Mustafa & Luiz Fernando Capretz

# Do Personality Profiles Differ in the Pakistani Software Industry and Academia – A Study

**Arif Raza**                                                                                       *arif_raza@mcs.edu.pk*
*Department of Computer Software Engineering,*
*National University of Sciences and Technology,*
*Islamabad, Pakistan.*

**Zaka-ul-Mustafa**                                                                                 *zaka@mcs.edu.pk*
*Department of Electrical Engineering,*
*National University of Sciences and Technology,*
*Islamabad, Pakistan.*

**Luiz Fernando Capretz**                                                                           *lcapretz@uwo.ca*
*Department of Electrical & Computer Engineering,*
*Western University,*
*London, Canada.*

**Abstract**

Effects of personality profiles and human factors in software engineering (SE) have been studied from different perspectives, such as: software life cycle, team performance, software quality attributes, and so on. This study intends to compare personality profiles of software engineers in academia and industry. In this survey we have collected personality profiles of software engineers from academia and the local industry in Pakistan. According to the Myers-Briggs Type Indicator (MBTI) instrument, the most prominent personality type among Pakistani academicians is a combination of Introversion, Sensing, Thinking, and Judging (ISTJ). However the most dominant personality type among software engineers in the Pakistani software industry is a combination of Extroversion, Sensing, Thinking, and Judging (ESTJ). The results of study establish that software engineers working in industry are mostly Extroverts as compared to those in academia who tend to be Introverts. The dimensions: Sensing, Thinking, and Judging (STJ), however, remain common in the dominant personality types of software engineers, both in Pakistani software industry and academia.

## 1. INTRODUCTION

Software life cycle traditionally involves different phases, such as requirement analysis, design, development, testing, and maintenance. People use their specific abilities to comprehend and carry out these activities successfully. Human factors and their personality traits play their role in each phase [1].

The Myers-Briggs Type Indicator (MBTI) [2] is primarily used for categorizing personality types. The indicator has four dimensions of preferences, which depict a particular personality. Within each facet, there are two pairs: Extroversion - Introversion, Sensing - Intuition, Feeling - Thinking, and Perceiving - Judging. Accordingly, sixteen typical personality dimensions are defined by combining these four distinctive types, each represented by four letters. Although an individual might have all eight preferences in each of the four dimensions, usually a person has one dominating dimension than its opposite side. The dimensions are further described in Appendix-A.

The sixteen types are used to describe people's personalities and temperaments according to MBTI. For example, if an individual is found to be the ENFJ type, it means that the individual prefers Extroversion, Intuition, Feeling, and Judging. This also indicates compatibility of

Arif Raza, Zaka-ul-Mustafa & Luiz Fernando Capretz

personality types with a particular career and how one makes decisions in life. It is possible though that these categories result in improved performance in specific situations, no category can be graded worse or better than the other.

MBTI has been variedly used to study human factors and personality profiles in different professions; software engineering has been one of them [3]. These studies include preferences and roles of human factors in different software life cycle phases, consequences of team work in software development, and effects and relationship of personality types on different tasks.

However, not much data is available in the South-Asian region about software professionals. A study has recently been carried out to acquire personality profiles and temperaments of software engineering faculty members and students in Pakistan [4].

The goal of this work is to identify and compare personality types of software engineering professionals in academia and industry. Accordingly, in this study, 18 professors of software engineering and 52 software engineers from Pakistani software industry are surveyed.

## 2. LITERATURE REVIEW

Sanz and Misra [5] observe that although software development depends a lot on human factors and personal performance, customarily functional and technical aspects of software gets more attention as compared to human and social factors. They maintain that software development is also affected from human factors like *"the role played by software professionals, the environment where they work, their qualification and skills, the motivation and attitude, etc."*

Boehm [6] suggests that software application should be integrated not only with hardware devices but with human controls as well. For software related research in future he emphasizes having a balance between software engineering, hardware and human factors.

Abduljalil and Kang [7] point out the gap in research related to human factors in software engineering. The authors consider human factors a must for software quality and human computer interaction. An empirical research model is also presented in the study called "*Intensive Prototype Model"* in order to enhance and facilitate the design process.

Cruz et al. [8] carry out a systematic literature review of 42 studies related to personality profiles in software engineering between 1970 and 2010. According to their findings, *"pair programming and team building are the most recurring research topics and MBTI is the most used test."*

Ahmed et al. [9] believe that there is a considerable impact of personality type over the way people become aware, make decisions and perform in different situations. In the empirical research, significance of personality types has been studied over the learning pattern of 85 software engineering students.

Varona et al. [10] study the personality profile of Cuban software engineers using the Myers-Briggs Type Indicator (MBTI). They find a combination of extroversion, sensing, thinking and judging as the most occurring personality type among the studied subjects.

Capretz and Ahmed [11] explore the need of personality diversity in the domain of software engineering with the aims of increasing consciousness in software engineers about importance of human factors, exploring effects of psychological factors to ward work efficiency, and to highlight significance of diversity in abilities and skills of software engineers. The authors look forward to more research on software psychology in the hope of understanding *"increased effectiveness and fulfillment among software engineers."*

According to Downey [12], Software Engineering (SE) students are generally not clear about their career path. The study is based on interviews of senior software professionals and points out the intricacies of software engineering profession.

Arif Raza, Zaka-ul-Mustafa & Luiz Fernando Capretz

Using MBTI, Cecil [13] studies personality profiles of IT professors and students in United States; it is concluded that association between students and professors' personality types plays an important role for students to continue their career in and IT.

Karn and Cowling [14] also make use of MBTI to study the impact of personalities on a software engineering (SE) team. The study depicts the use of ethnographic methods to comprehend human factors in SE. The study concludes that there exists an inclination of certain personality types towards certain roles.

Capretz [15] asserts that software engineers possess distinctive personalities. Although software engineering does have people of all psychological traits, certain types have dominant representation. According to this study, Introverts are more dominantly present in this field. They characteristically are not the masters of communication skills. This may explain why software systems are infamous for not addressing users' requirements.

Stanton [16] carries out two case studies to highlight the importance of human factors in interaction and interface design specification. The author states, *"human Factors sits between subject matter experts and software engineers, translating user requirements through the applications of theory, models and methods."*

Varona et al. [17] carry out a review of sixteen research studies in the area of human factors in software engineering between 1985 and 2011. The authors observe that due to the evolving changes in the software engineering processes, new challenges and roles have evolved for professionals as well.

## 3. RESEARCH METHODOLOGY

In this study, we surveyed 52 software engineers from the local industry and 18 software engineering professors from the academia. We visited four software houses; numerical distribution of the participants from industry is as follows: 14 Software Engineers took part in the survey from AERO (Advanced Engineering Research Organization, Hasan Abdal, Pakistan); 15 from Cyber Design, in Islamabad; 14 from PTCL (Pakistan Telecommunication Company Limited, Islamabad); and 17 software engineers from ID- Technologies, Islamabad. Conversely, from Academia, 18 professors of Software Engineering of the National University of Sciences and Technology, in Islamabad, Pakistan took part in the survey. A short version of the MBTI form (form G) was used to conduct the survey. The participation in the survey was on voluntarily basis, and no compensation in any form was offered to the participants.

Overall, the objective of this study is to answer the following research question: *"Whether software engineers working in industry and academia have different personality profiles?"*

## 4. RESULTS AND ANALYSIS

The personality type distribution of software engineers is summarized in Table 1 below. It can be observed that among the software engineers in academia (SEA), Introverts (72%) dominate over the Extroverts (28%), whereas among our respondents of software engineers in industry (SEI) Extroverts (52%) are more represented than Introverts (48%). Keeping in view the characteristics of Extroverts and Introverts, the observed data reflects that software engineers in industry focus more on the outer world and ideas, as compared to those in academia, who tend to focus on inner world of their own and prefer to work independently. Intuitive (56%) dominate over Sensing (44%) among SEA, whereas among SEI Sensing (58%) dominate over Intuitive (42%). This indicates that people in industry tend to focus on data and concrete facts, while academicians rely more on their imagination and inspiration. However, the results show the similarities in the rest of the two pairs of dimensions; both in SEA and SEI, Thinking are more popular than Feeling and Perceiving are less than Judging.

| I | E |
|---|---|
| SEA=72%<br>SEI=48% | SEA=28%<br>SEI=52% |
| **N**<br>SEA=56%<br>SEI=42% | **S**<br>SEA=44%<br>SEI=58% |
| **T**<br>SEA=67%<br>SEI=63% | **F**<br>SEA=33%<br>SEI=37% |
| **J**<br>SEA=56%<br>SEI=52% | **P**<br>SEA=44%<br>SEI=48% |

**TABLE 1:** Personality Type Distribution of Software Engineers in Academia (SEA) and Software Engineers in Industry (SEI)

Out of sixteen MBTI combinations, the ISTJ personality type has the top most representation of 22% among the surveyed Pakistani software engineering academicians, as shown in Table 2. This is followed by INTP with 17%, and then INFJ and ENFP both have 11%. ISFJ, INTJ, ISTP, ISFP, ENTP, ESTJ and ENTJ all have 5% representation. Among the respondents INFP, ESTP, ESFP, ESFJ, ENFJ have no representation. This is due to the small size of the sample.

| ISTJ | ISFJ | INFJ | INTJ |
|---|---|---|---|
| SEA =22%<br>SEI =12% | SEA =5%<br>SEI =4% | SEA =11%<br>SEI =2% | SEA =5%<br>SEI =2% |
| **ISTP**<br>SEA =5%<br>SEI =10% | **ISFP**<br>SEA =5%<br>SEI =4% | **INFP**<br>SEA =0 %<br>SEI =10% | **INTP**<br>SEA =17%<br>SEI =6% |
| **ESTP**<br>SEA =0 %<br>SEI =2% | **ESFP**<br>SEA =0 %<br>SEI =4% | **ENFP**<br>SEA =11%<br>SEI =6% | **ENTP**<br>SEA =5%<br>SEI =8% |
| **ESTJ**<br>SEA =5%<br>SEI =17% | **ESFJ**<br>SEA =0 %<br>SEI =6% | **ENFJ**<br>SEA =0 %<br>SEI =0% | **ENTJ**<br>SEA =5%<br>SEI =10% |

**TABLE 2:** The MBTI Types and their Distribution among SEA and SEI

ESTJ personality type has the top most representation (17%) among the surveyed Pakistani software engineers in industry. This is followed by ISTJ with 12% representation, ISTP, INFP and ENTJ having 10 % representation each as shown in Table 2. Among the respondents, ENFJ has got no representation at all in the sample.

## 5. DISCUSSION – IMPLICATIONS OF THE STUDY
Although there are many similarities in the type distribution of academia and industry, it is worth noticing that there are more ISTJ (22%) in academician sample and ESTJ (17%) in the industry sample, respectively, than any other type. The biggest discrepancies occur in the INTP and ESTJ cells, 17% in academia as opposed to 6% in industry, and 5% in academia against 17% in industry, respectively; the other remaining numbers for the other cells are more in tandem. It came as a surprise to find almost the same percentage of ISFJ and ISFP with 5% in academia and 4% in subjects from industry.

The study underscores the role of individual personality dimensions of the Pakistani Software Engineering community both in academia and in industry. According to the findings of the study, ISTJ configuration predominate in academia whereas ESTJ in the industry. INTP, INFJ and ENFP are the most dominant personality types in academia, whereas in industry, ISTJ, ISTP, INFP and ENTJ are the other well represented types. Overall, the results indicate that Introverts dominate in academia, whereas Extroverts lead in industry. Intuitive dominate over Sensing among SEA whereas Sensing dominate over Intuitive among SEI. Thinking participants outnumber Feeling and Judging dominates Perceiving in both the cases. The implications of these discrepancies highlight personality differences among software engineers in the software industry and in academia.

From a practical perspective, human resource departments of different employing companies could consider these personality traits while selecting software engineers. In this way, they could employ people with personality types that are a better fit to achieve the overall organizational objectives.

A constraint of this study is its limited number of respondents. Although the adopted approach has some potential threats to external validity, we followed apt research procedures to ensure the external validity. Moreover, in the surveys that we conducted, the respondents were neither influenced in any way nor were they offered any compensation.

## 6. CONCLUSION

The Software Engineering field seems to continuously attract people of diverse personality profiles and face the challenges that come along with this variation of personalities. According to the results and analysis of this study, although Extroverts dominate in industry and Introverts in academia, the other three dimensions Sensing, Thinking, and Judging (STJ) remain the dominant personality types of software engineers. In the future, we look forward for enhanced research in this important area and anticipate more reliable results with improved variance.

## 7. REFERENCES

[1]. L.F. Capretz, and F. Ahmed, (2010) "Making sense of software development and personality types," *IEEE IT Professional*, 12(1), 6-13.

[2]. I.B. Myers, M.H. McCaulley, N.L. Quenk, and A.L.Hammer, (1998) *MBTI manual - A guide to the development and use of the Myers-Briggs Type Indicator,* Palo Alto, California, Consulting Psychologists Press.

[3]. C.Bishop-Clark, and D.D. Wheeler, (1994) "The Myers Briggs personality type and its relationship to computer programming," *Journal of Research on Computing in Education,* 26(3), 358-370.

[4]. A. Raza, Z. ul-Mustafa, and L.F. Capretz, (2011) "Personality dimensions and temperaments of engineering professors and students – A survey," *Journal of Computing,* 3(11), 13-20.

[5]. L. Fernández-Sanz, and S. Misra, (2011) "Influence of human factors in software quality and productivity," *Computational Science and Its Applications - ICCSA Lecture Notes in Computer Science,* 6786/2011, 257-269.

[6]. B.W. Boehm, (2012) "Extending software engineering research outside the digital box," *Proceedings of the FSE/SDP workshop on Future of software engineering research,* ACM NY.

[7]. S. Abduljalil, and D.K. Kang, (2011) "Analysis of human factors in software application design for effective user experience," *Proceedings of the 13th International Conference on Advanced Communication Technology (ICACT),* 1446-1451.

[8]. S.S.J.O. Cruz, F.Q.B. da Silva, C.V.F. Monteiro, P. Santos, and I.Rossilei, (2011) "Personality in software engineering: Preliminary findings from a systematic literature review," *Proceedings of the 15th Annual Conference on Evaluation & Assessment in Software Engineering (EASE 2011),* 1-10.

[9]. F. Ahmed, P. Campbell, A. Jaffar, and S. Alkobaisi, (2010) "Innovations in practice learning & personality types: A case study of a software design course," *Journal of Information Technology Education,* 9.

[10]. D. Varona, L.F. Capretz, and Y. Piñero (2011) "Personality types of Cuban software developers," Global Journal of Engineering Education, 13 (2).

[11]. L.F. Capretz, and F. Ahmed, (2010) "Why do we need personality diversity in software engineering?" *ACM SIGSOFT Software Engineering*, 35 (2), ACM New York, USA.

[12]. J. Downey, (2010) "Careers in software: is there life after programming?" *Proceedings of the 2010 Special Interest Group on Management Information System's 48th annual conference on computer personal research,* ACM, NY, USA.

[13]. D.K. Cecil, (2009) "Personality types of IT professors," *ACM SIGITE, Fairfax, Virginia*, 13-23.

[14]. J.S. Karn, A.J. Cowling, (2006) "Using ethnographic methods to carry out human factors research in software engineering," *Behavior Research Methods,* 38(3), 495-503.

[15]. L.F. Capretz (2003) "Personality types in software engineering," *Int. J. Human-Computer Studies,* 58(2), 207-214.

[16]. N. Stanton (2012) "Human factors engineering as the methodological babel fish: translating user needs into software design," Human-Centered Software Engineering, 1-17.

[17]. D. Varona, L.F. Capretz, Y. Pinero, and A.Raza, (2012) "Evolution of software engineers' personality profile," ACM SIGSOFT Software Engineering Notes (SEN), 37(1), 1-5.

Arif Raza, Zaka-ul-Mustafa & Luiz Fernando Capretz

**Appendix-A**

**The Myers-Briggs Type Indicator (MBTI) Scales**

<u>Extroversion (E) versus Introversion (I):</u> Extroverts focus is on the outer world and ideas, like to communicate and mingle with other people, whereas Introverts tend to focus on inner world of their own and prefer to work independently.

<u>Sensing (S) versus Intuition (N):</u> Sensing individuals rely on facts and data they know, while Intuitive individuals use more frequently their imagination and inspiration.

<u>Thinking (T) versus Feeling (F):</u> Thinking people are cool-minded people who have logically defined tasks in their lives. Their decisions are based on logic and analytical rationale. Feeling people, on the other hand, are considered warm hearted, prefer to have pleasant working relationship and carry a sensitive approach.

<u>Judging (J) versus Perceiving (P):</u> Judging type prefers to have a plan, like to have stability in life, whereas Perceiving type keeps their options open to change, like spontaneity and remain flexible.